\documentclass[twocolumn,secnumarabic,amssymb, nobibnotes, aps, prl,showpacs]{revtex4}
\usepackage{amsmath,amssymb}
\usepackage{graphicx}

\begin{document}
\title{Spatio-temporal anomalous diffusion in heterogeneous media by NMR}
\author{Palombo M.$^1$ $^2$, Gabrielli A.$^1$ $^3$, De Santis S.$^1$ $^2$, Cametti C.$^1$, Ruocco G.$^1$ $^2$, Capuani S.$^1$ $^2$}
\email[Corresponding author: ]{silvia.capuani@roma1.infn.it}
\affiliation{$^{1}$Physics Department, Sapienza University of Rome, P.le A.Moro,5 00185 Rome, Italy\\
$^{2}$CNR IPCF UOS Roma, Physics Department, Sapienza University of Rome, P.le A.Moro,5 00185 Rome, Italy\\
$^{3}$ISC-CNR, Via dei Taurini, 19 00185 Rome, Italy}
\begin{abstract}
For the first time, the diffusion phase diagram in highly confined colloidal systems, predicted by Continuous Time Random Walk (CTRW), is experimentally obtained. Temporal and spatial fractional exponents, $\alpha$ and $\mu$, introduced within the framework of CTRW, are simultaneously measured by Pulse Field Gradient Nuclear Magnetic Resonance technique in samples of micro-beads dispersed in water. We find that $\alpha$ depends on the disorder degree of the system. Conversely, $\mu$ depends on both bead sizes and magnetic susceptibility differences within samples. Our findings fully match the CTRW predictions.
\end{abstract}
\pacs{82.56.Lz, 82.56.-b, 47.57.-s, 87.64.kj, 61.05.Qr, 76.60.-k}
\maketitle
\begin{figure}[t!]
\includegraphics[scale=.28]{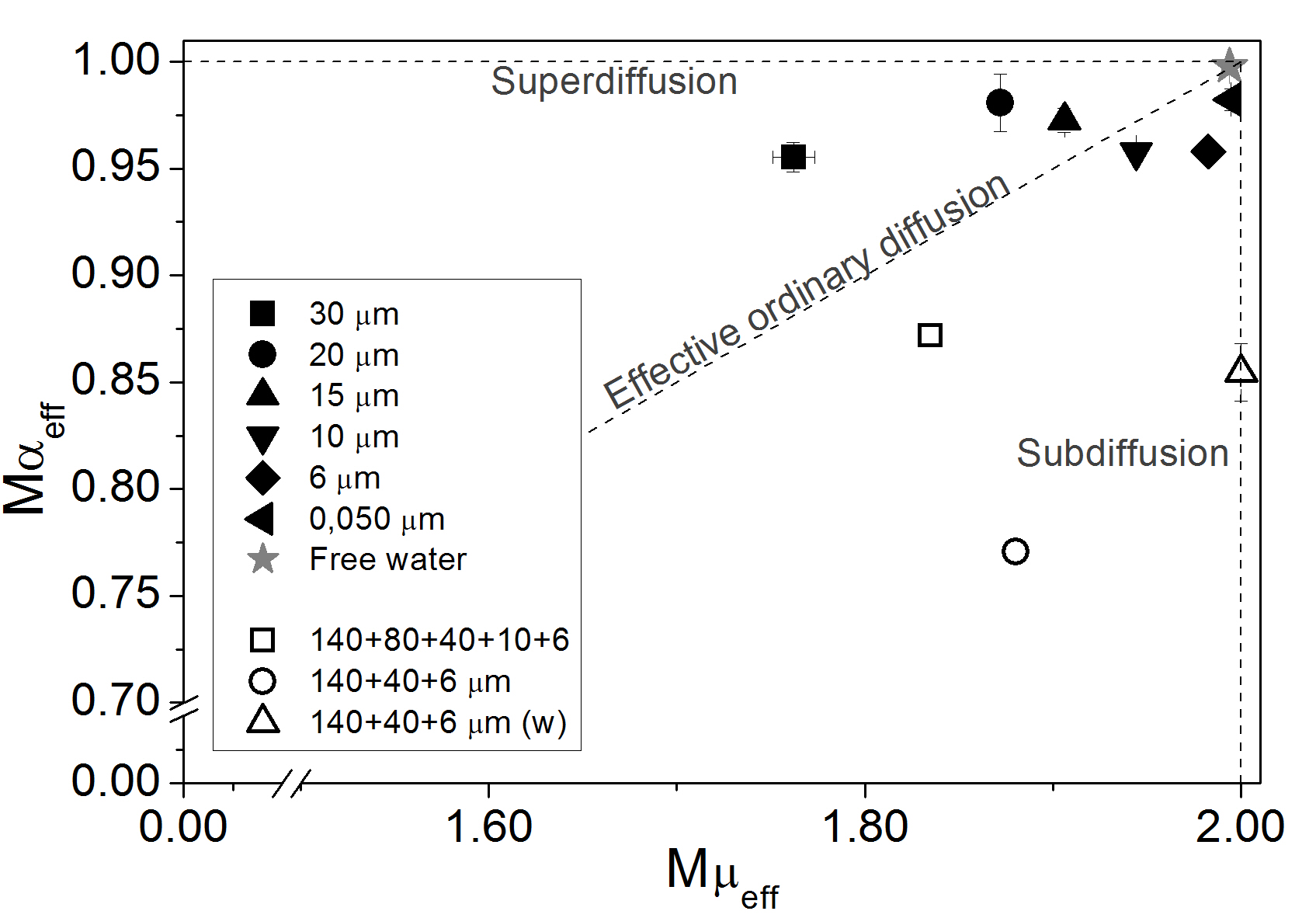}
\caption{$M\alpha_{eff}$ vs $M\mu_{eff}$ diagram for ordered samples (black filled symbols), disordered samples (empty symbols) and free water (star symbol). Dashed line $2M\alpha_{eff}=M\mu_{eff}$ represents the effective ordinary diffusion region in which $\langle x(t)^2\rangle\propto t^{\tfrac{2\alpha}{\mu}}=t$. The two dashed lines $M\alpha_{eff}=1$ and $M\mu_{eff}=2$ delimit the two regions, representing the superdiffusion ($\tfrac{2M\alpha_{eff}}{M\mu_{eff}}>1$) and subdiffusion ($0<\tfrac{2M\alpha}{M\mu}<1$) regimes.}
\label{fig:1}
\end{figure}
\begin{figure}[t!]
\includegraphics[scale=.24]{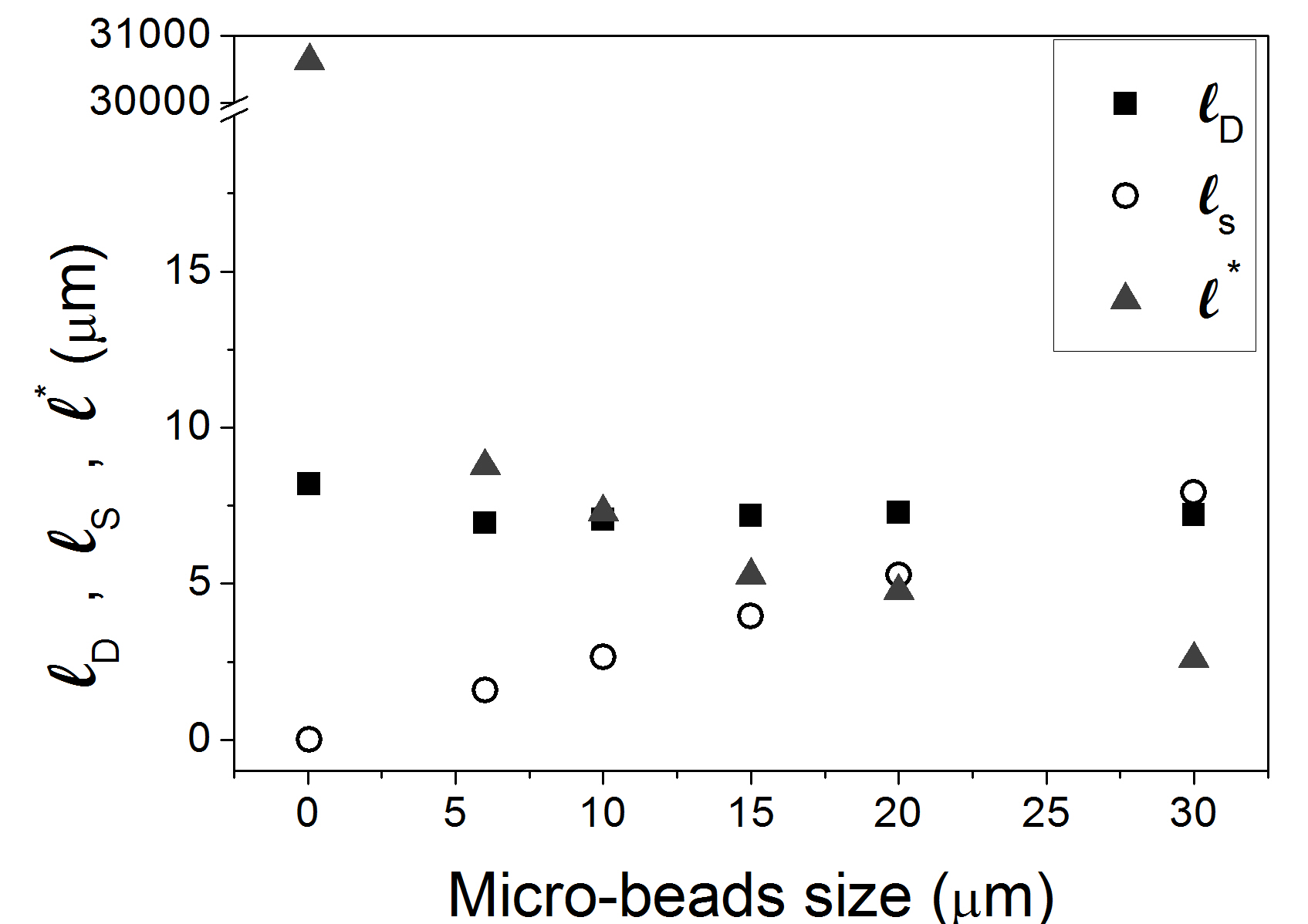}
\caption{Characteristic lengths $\ell_D$, $\ell_s$ and $\ell^*$ as a function of micro-beads size.}
\label{fig:2}
\end{figure}
\begin{figure}[t!]
\includegraphics[scale=.20]{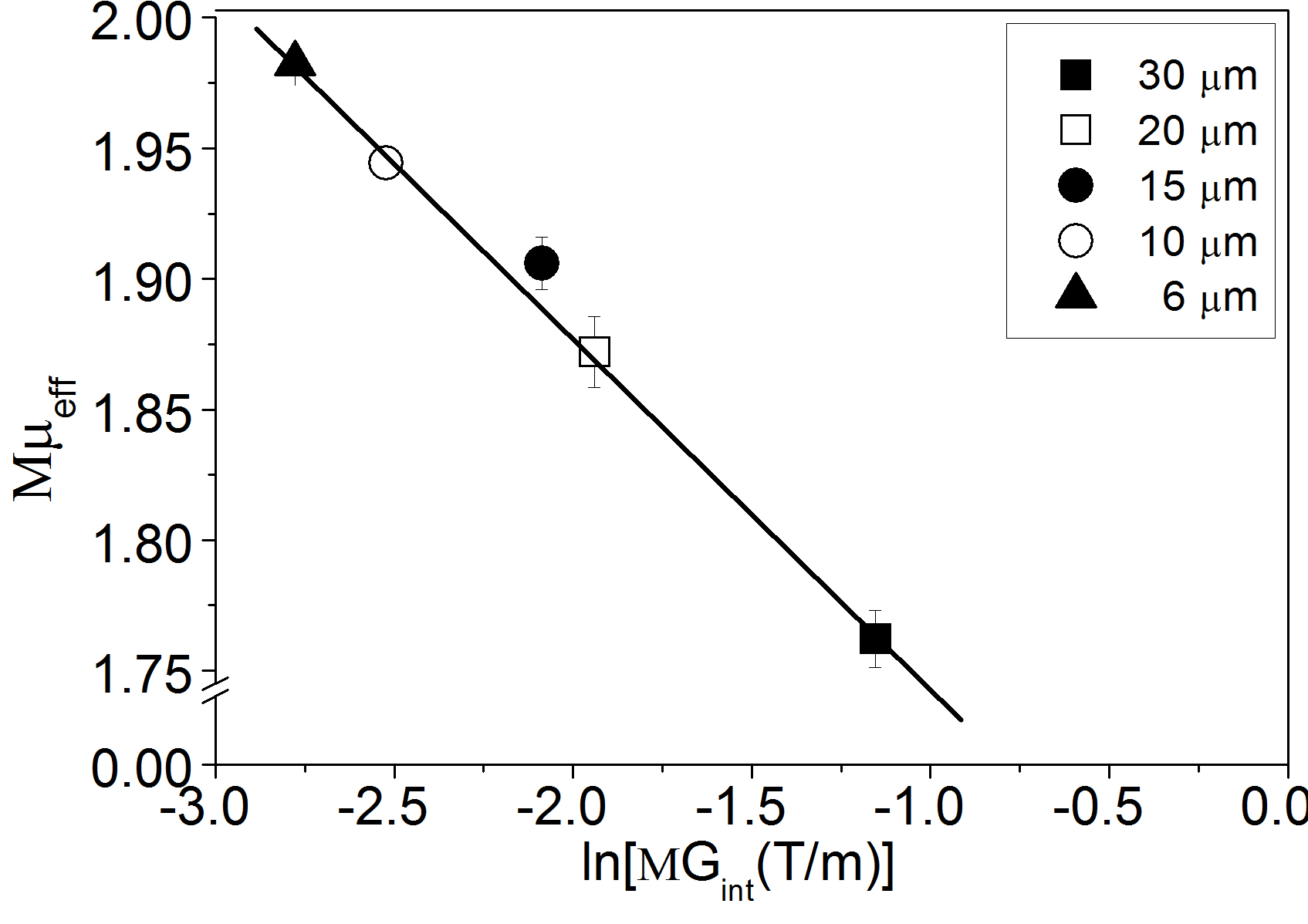}
\caption{$M\mu_{eff}$ as a function of $MG_{int}$, measured by PGSTE NMR sequence. Solid line is the regression line $M\mu_{eff}=(-0.1331\pm0.0061)\ ln(MG_{int}) + (1.613\pm0.016)$, $R=0.996$.}
\label{fig:3}
\end{figure}
\begin{figure}[t!]
\includegraphics[scale=.20]{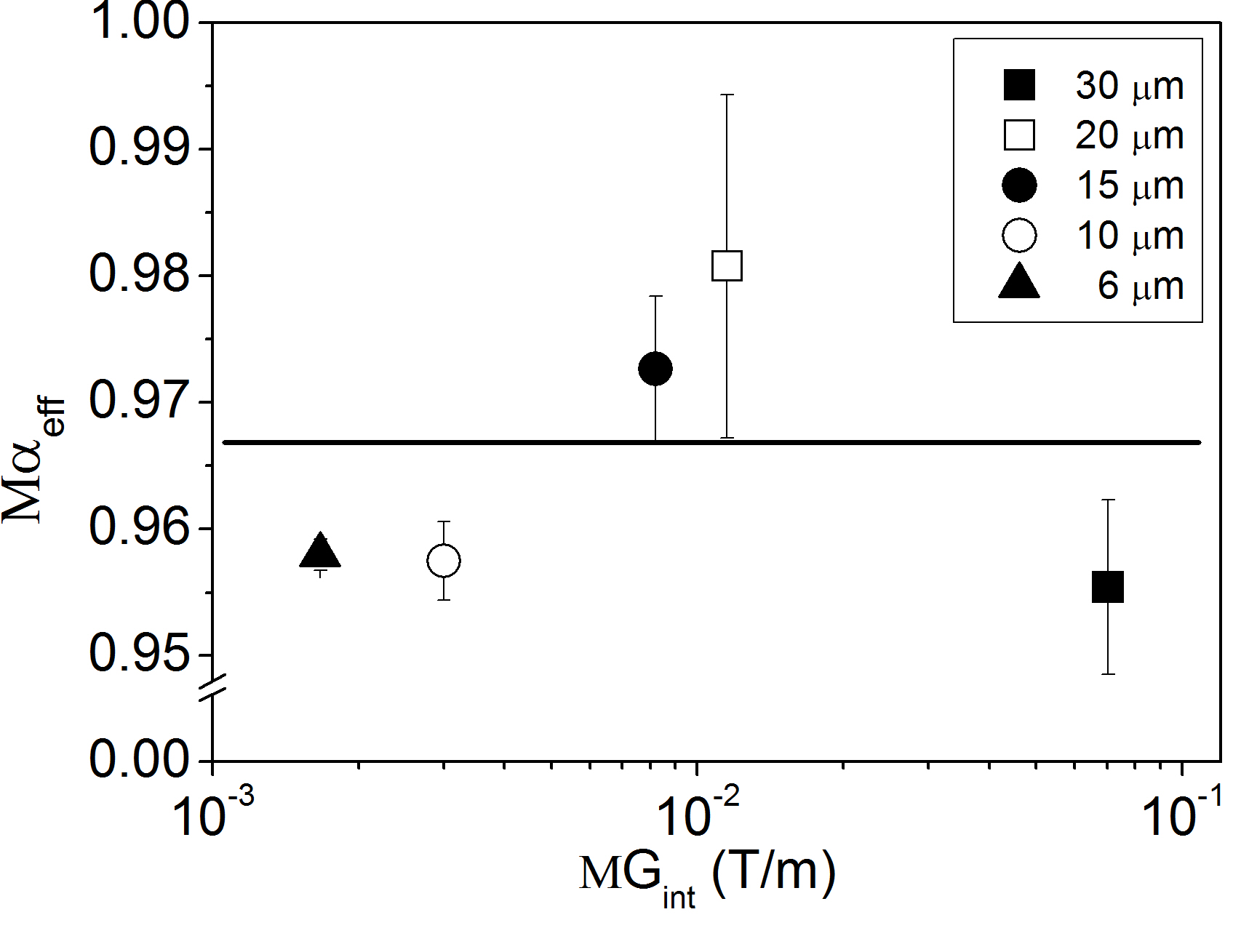}
\caption{$M\alpha_{eff}$ as a function of $MG_{int}$, measured by PGSTE NMR sequence. Solid line is the regression line $M\alpha_{eff}=(0.000\pm0.010)\ MG_{int} + (0.967\pm0.022)$, $R=0.051$.}
\label{fig:4}
\end{figure}

Anomalous diffusion (AD), that predicts the mean-square displacement (MSD) of a diffusing particle to grow nonlinearly in time, $\langle[x(t)-x(0)]^2\rangle\propto t^\nu$ (with $\nu\neq1$), is a property of many complex systems and its related phenomena have been observed in various physical fields \cite{Klemm,Solomon,Fischer,Plerou,Seisenberger}. As shown by Metzler and coworkers \cite{Metzler}, the features of AD can be described using fractional calculus. According to these Authors, molecular AD in media can be described by defining the motion propagator (MP) as the solution of fractional diffusion equations, which arise from the continuous time random walk (CTRW) model. These equations involve two fractional exponents, $\alpha$ and $\mu$, which are, the orders of the time and space fractional derivatives, respectively \cite{Schneider}. The theoretical framework of CTRW is well established and has been corroborated by huge amounts of Monte Carlo simulations (see for example \cite{Zoia}) together with several experimental evidences, mainly obtained by using fluorescent spectroscopy \cite{Periasamy,Wu}. However, to our knowledge, an experimental $\alpha$ \emph{vs} $\mu$ phase diagram, showing the competition between superdiffusion and subdiffusion of diffusing molecules in heterogeneous media, has never been carried out.\\
In this Letter, we experimentally challenge the Metzler et al. prediction of the CTRW theory \cite{Metzler} by measuring the fractional exponents $\alpha$ and $\mu$ by means of Nuclear Magnetic Resonance (NMR) methods, based on Pulse Field Gradient (PFG) technique \cite{Stejskal}, and by providing their interplay for the first time.\\
In the last few decades, NMR water diffusion measurements have been a topic of extensive research, having broad applications in biophysics \cite{Basser} and medicine \cite{Huang}. Translational self-diffusion in liquid systems can be measured using PFG techniques, by applying magnetic gradient pulses (named diffusion gradients) to the system, in addition to the static magnetic field of the instrument itself. The signal attenuation, that depends on both diffusion gradient strength and diffusion time, is simply the Fourier Transform (FT) of the MP. When MP is Gaussian, NMR signal attenuation follows a mono-exponential Stejskal-Tanner decay \cite{Stejskal}. Conversely, when the motion is described by a non-Gaussian propagator, the signal attenuation deviates from a mono-exponential decay \cite{Kopf,Ozarslan}.\\
In the present study, we have measured the fractional exponent $\alpha$ collecting the PFG signal attenuation as a function of diffusion times compared to the asymptotic expression of FT of the MP for subdiffusive regime, obtained from \cite{Metzler}. Moreover, $\mu$ was extracted from the PFG signal attenuation as a function of gradients strength fitted to the expression of FT of the MP for superdiffusive regime, obtained again from \cite{Metzler}. PFG NMR experiments were performed on controlled porous media characterized by packed polystyrene micro-beads (Microbeads AS, Norway) of various sizes dispersed in water. Mono-dispersed (mono-d) and poly-dispersed (poly-d) samples were used to check the potential ability of $\alpha$ and $\mu$ parameters to detect different degrees of system disorder. Moreover, internal magnetic field gradient ($G_{int}$) measurements were performed to investigate the influence of the magnetic susceptibility differences ($\Delta\chi_m$) on measured $\alpha$ and $\mu$ parameters. Experimental results demonstrate that the exponent $\alpha$ strongly depends on the disorder degree of the system, and is independent of $\Delta\chi_m$. Conversely, the exponent $\mu$ is strongly correlated with $\Delta\chi_m$ and pore sizes. In summary, mono-d samples are characterized by both ordinary diffusion and spurious magnetic susceptibility effect, while poly-d samples show a real subdiffusion of water molecules. Thanks to peculiar features of NMR technique, an experimental evidence of the spatio-temporal competition between long rests and long jumps of diffusing molecules, as defined in \cite{Metzler}, is reported here.
\paragraph{Theory}Taking into account the CTRW model \cite{Metzler}, each displacement in a diffusion process can be characterized by two parameters: a waiting time $\tau$ elapsing between two consecutive steps, and a variable displacement length $\xi$. $\tau$ and $\xi$ are independent random variables, distributed according to the probability densities $\psi(\tau)$ and $\lambda(\xi)$.
When $\psi(\tau)\simeq \tau^{-1-\sigma}$, with $0<\sigma<1$, and $\lambda(\xi)$ is characterized by a finite variance, the characteristic waiting time $T=\int_0^\infty \tau\psi(\tau)d\tau$ diverges and the resulting CTRW is \emph{subdiffusive}. As a consequence, the asymptotic behaviors of the FT of the MP is
\begin{equation}
\label{alphaLk}
W(k,t)\simeq
\left\{
\begin{aligned}
e^{-K_\alpha k^2 t^\alpha}\quad \text{when $k^2\ll \tfrac{1}{K_\alpha t^\alpha}$}\\
\frac{1}{K_\alpha k^2 t^\alpha}\quad \text{when $k^2\gg \tfrac{1}{K_\alpha t^\alpha}$}
\end{aligned}
\right.
\end{equation}
where $\alpha=\sigma$ and, if $\sigma\geq1$, $\alpha=1$.
Under these conditions, the MSD grows sublinearly in time, namely $\langle x^2(t)\rangle \simeq t^{\alpha}$.
\emph{Vice-versa}, when $\lambda(\xi)\simeq |\xi|^{-1-\rho}$, with $0<\rho<2$, and $\psi(\tau)$ is characterized by a finite mean value, the jump length variance $\Sigma^2=\int_{-\infty}^{\infty} \xi^2\lambda(\xi)d\xi$ diverges and a \emph{superdiffusive} (Levy flight) CTRW is obtained. In this case, the behavior of the FT of the MP is
\begin{equation}
\label{mu}
W(k,t)\simeq e^{-K^\mu|k|^\mu t}
\end{equation}
where $\mu=\rho$ and, if $\rho\geq2$, $\mu=2$.
In the case $\alpha<1$ and $\mu<2$, when spatio-temporal coupling of jump length and waiting time is taken into consideration, an $\alpha$ \emph{vs} $\mu$ phase diagram can be drawn, as reported by \cite{Metzler}. In the specific case shown here, i.e., mono-d and poly-d micro bead samples, we can expect to detect subdiffusion (i.e. characterized by $\alpha<1$ and $\mu=2$) and Gaussian (i.e characterized by $\alpha=1$ and $\mu=2$) diffusion processes only.\\
In NMR spectroscopy, the effect of diffusion on an \emph{ensemble} of $N$ magnetized spins can be described in terms of the behavior of spin phases, $[\Phi(\mathbf{r}_i)]_{i=1...N}$, subjected to an applied magnetic field gradient \cite{Callaghan}. The PFG NMR technique is characterized by radio frequency pulses together with a couple of magnetic field gradient pulses of duration $\delta$, strength $|\mathbf{g}|$, and separated by a delay time $t=\Delta$.
The first gradient pulse effectively phase encodes the molecular spins according to each molecule position. The second decoding gradient pulse completely  reverses the spin phase if and only if the molecule has not diffused during the time $\Delta$. Conversely, if the molecule diffuses, the intensity of the signal due to the molecule will be attenuated because of a non complete phase reverse. In case of ordinary diffusion, the degree of attenuation is a function of both $\mathbf{g}$ and $\Delta$, and occurs at a rate proportional to the self-diffusion coefficient $D$. One of the most interesting features of PFG is that the measured signal, $E_\Delta(\mathbf{k})$, is the FT of the averaged propagator, $\overline{P}_s(\mathbf{R,\Delta})$, i.e. $E_\Delta(\mathbf{k})\propto W(\mathbf{k},\Delta)$.
Since $\Delta$ defines the time window in which the diffusion processes are observed, PFG NMR techniques allow the measure of \emph{effective} $\alpha$ and $\mu$ values ($\alpha_{eff}$ and $\mu_{eff}$) using the aforementioned expressions \eqref{alphaLk} and \eqref{mu}, respectively. In particular, $\alpha_{eff}$ is obtained by a fitting procedure of experimental data measured by PFG at different $\Delta$ values to the expression \eqref{alphaLk}. The coefficient $\mu_{eff}$ is instead obtained by a fitting procedure of experimental data at different $\mathbf{k}=\tfrac{1}{2\pi}\gamma \mathbf{g} \delta$ values, where $\gamma$ is the nuclear gyromagnetic ratio. Taking into account the $\Delta$ values usually selected to perform PFG experiments, we expect to measure $\mu_{eff}=2$ and $\alpha_{eff}\leq1$, depending on the investigated heterogeneous sample.
Another peculiar property of the NMR signal in heterogeneous systems is its dependence on $\Delta\chi_m$ generated at the interface between regions with different $\chi_m$. The presence of different susceptibilities can be quantified through $G_{int}$ measurements, which can be extracted by the signal decay, generated by a Spin Echo (SE) NMR sequence \cite{Callaghan,GradInt}. As a consequence, both $\alpha_{eff}$ and $\mu_{eff}$ could be, in principle, affected by $\Delta\chi_m$.
\paragraph{Methods and Materials}All measurements were performed on a Bruker $9.4\ T$ Avance system, operating with a micro-imaging probe ($10\ mm$ internal diameter bore) and equipped with a gradient unit characterized by a maximum gradient strength of $1200\ mT/m$, and a rise time of $100\ \mu s$. As a first step, relaxation-time NMR measurements were performed to fully characterize the investigated porous samples. In order to determine the porosity $p$, as previously reported \cite{Amitay-Rosen}, a SE imaging version using MSME (Multi Slice Multi Echo) sequence (repetition time $TR=1500\ ms$, matrix $128\times128$, slice thickness $STH=1\ mm$, in plane pixel dimension $60\times60\ \mu m^2$, number of scan $NS=8$) at various echo times $TE$ (from $2.8\ ms$ to $300\ ms$) was used to obtain SE decay in different regions of each sample. Conventional mean diffusion coefficient $\overline{D}$ of water in each sample was measured by means of a spectroscopic Pulsed Field Gradient STimulated Echo (PGSTE, $TE/TR=18/3000\ ms$, diffusion gradient pulses delay $\Delta=80\ ms$, diffusion gradient pulses duration $\delta=4.4\ ms$ and diffusion gradient strength $g$ applied along the $x$, $y$ and $z$ axes using $32$ gradient amplitude steps from $6\ mT/m$ to $100\ mT/m$). $G_{int}$ along $x$, $y$ and $z$ axes was measured by using a spectroscopic SE sequence ($TR=1500\ ms$, $NS=8$) with $N=64$ data points (corresponding to $64$ echoes refocusing every $2\ ms$ from $1\ ms$ to $125\ ms$), as previously described \cite{DeSantis}. Mean value of internal magnetic field gradient ($MG_{int}$) was obtained as the average value along $x$, $y$ and $z$ axes. The characteristic diffusion length $\ell_D=(2\overline{D}\Delta)^{1/2}$ and the characteristic length $\ell^*=(\tfrac{\overline{D}}{\gamma G_{int}})^{1/3}$, as defined in \cite{GradInt}, were obtained from these preliminary measurements. In particular, $\ell^*$ establishes the spatial region around each bead, within which internal gradients act, dephasing the spins.
A spectroscopic PGSTE with $\delta=4.4\ ms$, $g=0.10\ T/m$ (i.e. $k=22481\ m^{-1}$) along $x$, $y$ and $z$ axes, with $TR=2.5\ s$, $NS=32$ and $48$ values of $\Delta$ in the range $(0.020\div1.0)\ s$ was used to collect data to be fitted to equation \eqref{alphaLk}, in order to extract the $\alpha_{eff}$ value along $x$, $y$ and $z$ axes. \emph{Vice-versa}, a spectroscopic PGSTE with $\Delta/\delta=80/4.4\ ms$, $TR=2.5\ s$, $NS=16$ and $48$ gradient amplitude steps from $0.026$ to $1.02\ T/m$ along $x$, $y$ and $z$ axes was used to collect data to be fitted to the function \eqref{mu}, thus obtaining a measure of the $\mu_{eff}$ value along $x$, $y$ and $z$ axes. The mean values of $\alpha_{eff}$ ($M\alpha_{eff}$) and $\mu_{eff}$ ($M\mu_{eff}$) were obtained by averaging: $M\alpha_{eff}=\tfrac{1}{3}\sum_{i=x,y,z}(\alpha_{eff})_i$ and $M\mu_{eff}=\tfrac{1}{3}\sum_{i=x,y,z}(\mu_{eff})_i$.\\
All fitting procedures were performed by means of Levenberg-Marquardt algorithm.\\
Nine samples in total were carried out using polystyrene micro-beads with nominal average diameters of  $30$, $20$, $15$, $10$, $6$ and $0.050\ \mu m$ and characterized by a $|\Delta\chi_m|=|\chi_m^{H_2O}-\chi_m^{polystyrene}|\sim 1.59$ in SI units. Six $10\ mm$ NMR tubes were filled up to a volume of approximately $2\ cm^3$ with beads mono-d in a solution of polyoxyethyle-sorbitan-mono-laurat (Tween 20) $10^{-6}\ M$ and deionized water (conductivity$\sim10^{-6}\ mho/cm$), and investigated $1$ month after their preparation. The sphere packing in each mono-d sample, was characterized by a mean "sphere density" $\eta=0.72\pm0.03$. As a consequence, water diffused in porous systems comprised by interconnected pores of mean diameter $\ell_s$ greater than the one of cannon-ball packing configuration, i.e. $\ell_s\geq\tfrac{\sqrt{3}+1}{4}d$, where $d$ is the bead mean diameter.
Three other poly-d specimens were prepared in a similar way, using equal volume fractions of beads of varying sizes. The sample called $140+80+40+10+6\ \mu m$, was made with micro-beads of $140$, $80$, $40$, $10$ and $6\ \mu m$,  while two other samples called $140+40+6\ \mu m\ (w)$ and $140+40+6\ \mu m$ were made with micro-beads of $140$, $40$ and $6\ \mu m$ investigated $1$ week (w) and $1$ month (m) after their preparation, respectively. The latter sample ($\eta=0.41\pm0.03$, $p=0.59\pm0.03$) is the most disordered one, while the former ($\eta=0.69\pm0.03$, $p=0.31\pm0.03$) is less disordered when compared to the $140+40+6\ \mu m$ specimens, but more disordered when compared to mono-d samples. Finally, one tube filled up with free water was also used as control.
\paragraph{Results and discussion}The experimental version of the theoretical $\alpha$ \emph{versus} $\mu$ graph suggested by Metzler et al. \cite{Metzler}, for all ten samples investigated, is shown in FIG.\ref{fig:1}. Data points obtained from ordered mono-d micro-bead samples (black filled symbols) lie close to the dashed line $\alpha_{eff}=\tfrac{\mu_{eff}}{2}$, indicating the border line between superdiffusion and subdiffusion regime zones. These points are identified by a similar $M\alpha_{eff}$ value very close to $1$, and by different values of $M\mu_{eff}$, ranging from $1.75$ to $2$, depending on bead sizes. Specifically, data points associated to $0.050\ \mu m$, $6\ \mu m$ and $10\ \mu m$ mono-d samples are localized in a subdiffusion regime zone, while $15\ \mu m$, $20\ \mu m$ and $30\ \mu m$ ones lie in the superdiffusion regime zone, close to the border line. The distribution of these data matches the experimental results showed in FIG.\ref{fig:2}, where the dynamic feature of water in all mono-d samples is described by the interplay between the characteristic lengths $\ell_D$, $\ell_s$ and $\ell^*$ \cite{GradInt}. When $\ell_s<\ell_D,\ell^*$ (i.e. bead size lower than $15\ \mu m$ in the investigated samples), motional averaging regime occurs and spins explore the entire pore many times before a dephasing arises, so any local magnetic field variation is averaged out by diffusion. Conversely, when $\ell^*<\ell_D,\ell_s$ (i.e. bead size is equal to and higher than $15\ \mu m$ in the investigated samples), the slow diffusion case occurs and spins experience a no totally averaged $G_{int}$, which produces an additional spin dephasing \cite{GradInt}.
\emph{Vice-versa}, data points obtained from disordered poly-d micro-bead samples (empty symbols) lie, as expected, in the subdiffusive region. Specifically, $M\alpha_{eff}$ decreases as the degree of the disorder increases, while $M\mu_{eff}$ does not discriminate between ordered and disordered samples, and assumes values slightly smaller than $2$. The broader the distribution of spatial length scales which characterize poly-d samples, the broader the distribution of characteristic resting time scales. As a consequence, water diffusing in investigated poly-d samples has a non negligible probability to experience long resting times and, thus, to show an anomalous subdiffusive behavior quantified by the decreasing $M\alpha_{eff}$ value from the unity, as bead disorder increases.
To understand the physical meaning of the $\mu_{eff}$ parameter, which in principle accounts for superdiffusion processes, we guessed that the dependence of $\mu_{eff}$ on bead size may be caused by $\Delta\chi_m$ at the interface between beads and diffusing water. To validate this assumption, we obtained the plot of FIG.\ref{fig:3}, which shows a strong linear correlation ($R=0.996$) between $M\mu_{eff}$ and $ln(MG_{int})$ for all mono-d samples. In particular, the higher the $MG_{int}$, the lower the $M\mu_{eff}$ value. Conversely $M\alpha_{eff}$ does not depend on $\Delta\chi_m$ quantified by $MG_{int}$, as shown in FIG.\ref{fig:4}. In the light of these correlations, it is reasonable to assume that values of $M\mu_{eff}$ less than $2$ are spuriously due to $\Delta\chi_m$, that introduces a pseudo-absorb/desorb process of water molecules at the interface between bead surface and water, instead of a real superdiffusion mechanism. This is also shown in FIG.\ref{fig:2}, which clearly shows that slow diffusion processes become predominant as pore diameter $\ell_s$ increases. In particular, local gradients impart a phase shift to the spins within $\ell^*$ region that adds up to the phase shift given by the diffusion gradient pulse.
As a consequence, some spins contribute to increase the degree of PFG signal attenuation; differently, other spins, which can be in a very distant zone from the first ones, will acquire a phase which will help to increase the signal. Due to indistinguishable spins associated to water molecules, this scenario simulates a superdiffusion regime of water molecules whose signal disappears in one spot and appears in another. In other words, we suggest that in the investigated micro-bead samples, the measured $\mu_{eff}$ values less than $2$ are only due to artifact effects generated by $\Delta\chi_m$ at the interface between beads and diffusing water. Indeed it is known that water behavior in these kind of interconnected micro-pore systems is Brownian or subdiffusive \cite{Zoia,Horton}. However, these artifacts, which simulate superdiffusion, provide information about different pore sizes. These observations give a basis for further investigations on the correlation between $\mu_{eff}$ and $G_{int}$.
\paragraph{Conclusion}Here we have measured the characteristic parameters of diffusion phase diagram, $\alpha$ and $\mu$, by means of NMR in systems marked by the presence of spatio-temporal competition between long rests and long jumps of diffusing molecules. Following the basic common knowledge on complex systems, we made ordered and disordered media using mono- and poly-dispersed micro-beads. The NMR measure of their disorder degree fully matches the predictions. Moreover, we highlighted that the real physical mechanism, which gives rise to an apparent superdiffusion, is due to $\Delta\chi_m$ between the solid and liquid phase in heterogeneous samples. Inferring disorder features in heterogeneous media from diffusion phase diagram obtained by NMR may be of paramount value in a variety of applications from oil-well logging and dynamics of polymers to the diagnosis and monitoring in vivo of many diseases in the human body.
\bibliographystyle{plain}
\bibliography{bibliography_2}
\end{document}